%% file: main.tex
\documentclass{article}
\usepackage[T1]{fontenc} 
\usepackage[utf8]{inputenc} 
\usepackage{ismir,amsmath,cite,url}
\usepackage[caption=false]{subfig}
\usepackage{graphicx}

\usepackage{soul}
\usepackage{ulem}
\usepackage{array}

\newcolumntype{M}[1]{>{\centering\arraybackslash}p{#1}}

\MakeRobust{\subref}

\newsavebox{\leftimage}
\newsavebox{\rightimage}
\newlength{\imageheight}

\usepackage{color}
\usepackage{algorithm}
\usepackage{algpseudocode}
\usepackage{subcaption}
\usepackage{multirow}

\usepackage{lineno}
\usepackage{subfiles}

\usepackage{enumitem}

\setlist[itemize]{itemsep=2pt, parsep=2pt, topsep=2pt, partopsep=2pt,leftmargin=*,labelsep=1em}

\setlist[enumerate]{itemsep=2pt, parsep=2pt, topsep=2pt, partopsep=2pt,leftmargin=*,labelsep=1em}

\title{SymPAC: Scalable Symbolic Music Generation With Prompts And Constraints}

\multauthor{Haonan Chen$^1$ \hspace{0.4cm} Jordan B. L. Smith$^2$ \hspace{0.4cm} Janne Spijkervet$^1$ \hspace{0.4cm} Ju-Chiang Wang$^1$} 
{ \bfseries{Pei Zou$^1$ \hspace{1cm} Bochen Li$^1$ \hspace{1cm} Qiuqiang Kong$^3$ \hspace{1cm} Xingjian Du$^1$}\\
 $^1$ ByteDance Inc. San Jose, USA \\
 $^2$ Queen Mary University of London \\
$^3$ Department of Electronic Engineering, The Chinese University of Hong Kong \\
{\tt\small haonanchen@bytedance.com}
}

\def\authorname{H. Chen, J. BL Smith, J. Spijkervet, J.-C. Wang, P. Zou, B. Li, Q. Kong, and X. Du}

\begin{document}

\maketitle

\begin{abstract}
Progress in the task of symbolic music generation may be lagging behind other tasks like audio and text generation, in part because of the scarcity of symbolic training data.
In this paper, we leverage the greater scale of audio music data
by applying pre-trained MIR models (for transcription, beat tracking, structure analysis, etc.) to extract symbolic events and encode them into token sequences.
To the best of our knowledge, this work is the first to demonstrate the feasibility of training symbolic generation models solely from auto-transcribed audio data.
Furthermore, to enhance the controllability of the trained model, we introduce SymPAC (\textbf{Sym}bolic Music Language Model with \textbf{P}rompting \textbf{A}nd \textbf{C}onstrained Generation), which is
distinguished by using
(a)~\textit{prompt bars} in encoding and (b)~a technique called \textit{Constrained Generation via Finite State Machines (FSMs)} during inference time.
We show the flexibility and controllability of this approach,
which may be critical in making music AI useful to creators and users.

\end{abstract}

\section{Introduction}\label{sec:introduction}

The success of language models — especially large ones — has demonstrated that with
more data and larger models,
using a simple language model objective can endow a model with powerful natural language generation capabilities.
On the other hand, although symbolic music and natural language share many similarities,
no music model has yet seemed to match the capabilities of generative text models.
One reason for this gap is the insufficient amount of symbolic music data.

To address this, previous efforts in symbolic music generation have involved combining limited manually annotated data with data obtained by automatic transcription~\cite{choi2020encoding}, or collecting private symbolic training datasets~\cite{zeng2021musicbert}.
By contrast, in this work, we demonstrate that a high-quality, multi-track symbolic music generation model can be trained just using results from running Music Information Retrieval (MIR) models on audio music data.
In this way,
our framework eliminates the need for manually annotated symbolic music data, allowing for expansion purely through audio datasets.

On the other hand,
there has been a recent surge of efforts
that directly generate the auditory modality of music~\cite{musiclm,huang2023noise2music,copet2024simple}.
This is useful for some applications, but typically precludes fine-grained control and
editing the outcome,
which is crucial for composers who wish to shape their musical ideas precisely.
In contrast, outputting symbolic data gives composers the ability to interactively shape and modify their musical ideas.

Considering such advantages, the problem of how to integrate user input to control the generation of symbolic music has been a popular research topic. In previous works, two methods for incorporating control signals are usually used. The first approach is based on a Variational Autoencoder (VAE)~\cite{roberts2018hierarchical,wu2023musemorphose}, wherein the control is exerted within the VAE's latent space.
The second approach is to embed control information directly into the encoding of symbolic music and implant control inputs during inference \cite{musenet,ens2020mmm,von2022figaro,dong2023multitrack}.

In this work, we introduce the \textbf{SymPAC} framework (\textbf{Sym}bolic Music Language Model with \textbf{P}rompting \textbf{A}nd \textbf{C}onstrained Generation),
designed to work with decoder-only language models to enable user input controls.
The SymPAC framework consists of the following two parts.
First, inspired by the prompting mechanism used in the natural language domain \cite{radford2019language,brown2020language}, we introduce \textit{prompt bars} in our symbolic music encoding, which consolidates all control signals into a separate prompt section before encoding the actual musical notes.
This design is essential for a decoder-only language model to have the full context of control signals during the generation of music.
Second,
in the controlled symbolic music generation setting, the generated tokens should not only comply with the encoding grammar but also adhere to user inputs.
Thus we propose to use \textit{Constrained Generation via Finite State Machines (FSMs)}, which constrains the sampling of tokens at each time step to a subspace.
We will discuss the advantages of SymPAC over previous methods in Section~\ref{sec:related}, and provide more details of how SymPAC can be used for various types of user inputs in Sections~\ref{sec:method} and~\ref{sec:experiment}.

We collected roughly
one million in-house audio samples and extracted MIR information for each, using
pre-trained models for beat tracking \cite{hung2022modeling}, chord detection \cite{lu2021spectnt}, section detection \cite{wang2022catch,wang2022musfa}, multi-track transcription \cite{lu2023multitrack}, and music tagging \cite{won2021semi}.
The MIR results were transformed into various tokens, and then integrated into an extended REMI \cite{huang2020pop,von2022figaro} encoding to train a language model based on Llama \cite{touvron2023llama} architecture.
To summarize, our main contributions are:

\noindent\textbf{Scalability}: We demonstrate that a high-quality symbolic music generation model can be trained solely with transcribed data, without the need of manually annotated symbolic music, and can be scaled by amassing more audios.

\noindent\textbf{Controllability}: We propose the SymPAC framework, which enables flexible user input controls on a decoder-only language model while retaining good quality.

\begin{table*}[!ht]
 \begin{center}
 \begin{tabular}{llllcc}
  \hline
  \textbf{Dataset} & \textbf{\#Songs} & \textbf{\#Notes} & \textbf{Format} & \textbf{Multitrack} & \textbf{Public} \\
  \hline\hline
  Maestro \cite{hawthorne2018enabling} & 1.1K & 6M & MIDI & N & Y \\
  GiantMIDI-Piano \cite{kong2022giantmidi} & 10.9K & 39M & MIDI & N & Y \\
  Lakh \cite{lakhmidi} & 170K & 910M & MIDI & Y & Y \\
  MMD \cite{zeng2021musicbert} & 1.5M & 2,075M & MIDI & Y & N \\
  \hline
  FMA \cite{fma_dataset} & 100K & N/A~* & Audio & Y & Y \\
  MSD \cite{msd} & 1M & 709M & Audio & Y & Y \\
  DISCO-10M \cite{lanzendorfer2024disco} & 15M & N/A~* & Audio & Y & Y \\
  In-House Dataset (IHD) & 1M & 3,688M & Audio & Y & N \\
  \hline
 \end{tabular}
\end{center}
\centering
 \caption{Comparison of different symbolic and audio music datasets. *~Since we did not run transcription on FMA or DISCO-10M, we don't have the number of notes information for them.}
 \label{tab:dataset}
\end{table*}

\section{Related Work}\label{sec:related}
\subsection{Training Data For Symbolic Music}
In Table \ref{tab:dataset}, we summarize some popular music datasets in the symbolic and audio domains, together with our in-house audio dataset, and compare their sizes.
The Lakh MIDI Dataset \cite{lakhmidi} is one of the biggest public datasets, containing 170K multitrack pieces in MIDI format.
Many researchers use publicly available symbolic music datasets for training, but some collect and use large-scale ones that are not disclosed; e.g., MusicBERT~\cite{zeng2021musicbert} was trained on the Million-MIDI Dataset (MMD).

Although the combined size of the public datasets in Table~\ref{tab:dataset} is large, combining them is not straightforward since they vary in format.
For example, the Maestro dataset consists of transcriptions of piano performances where note timings reflect actual performance timings, whereas datasets like Lakh are quantized to metrical time with alignment to beats. The inclusion of instrument tracks and additional information (e.g., chords, sections) also differs between datasets.
To expand the scale of training data by combining these datasets, it is necessary to unify their formats first, which may be tedious and introduce errors. 

On the other hand, publicly available audio datasets are much larger in scale. The Million Song Dataset (MSD) \cite{msd}, for example, contains 1M songs, or 709M notes in total after being run through a 5-track transcription model \cite{lu2023multitrack}.
The recently published DISCO-10M \cite{lanzendorfer2024disco} is of an even larger scale.
Furthermore, by using a single set of MIR models to annotate all the audio data,
we do not need to be concerned about the issue of inconsistent data formats.
This makes it easier to scale up the training dataset.

\subsection{Encoding For Symbolic Music}
Since the introduction of the Music Transformer \cite{huang2018music}, language models based on the transformer architecture have become a popular choice for symbolic music generation.
One of the most critical research questions has been how to encode symbolic music that is amenable to processing by such a model, which, in the context of language models, involves converting the piece into a sequence of tokens.

Early transformer-based models for symbolic music predominantly employed a MIDI-like encoding scheme, by treating MIDI event sequences almost identically as input token sequences \cite{musenet,donahue2019lakhnes,ens2020mmm}.
Later, the Revamped MIDI (REMI) encoding \cite{huang2020pop} was proposed, which modified the MIDI encoding by replacing time shift events with duration events for each note and introducing bar and beat concepts to adopt metrical time instead of absolute time. These modifications facilitated the model's learning of rhythmic patterns within the music, 
improving the quality of the output.
Building upon REMI, several extensions have been proposed to support encoding multitrack \cite{ens2020mmm} and various control tokens \cite{von2022figaro}.
Our work is based on the multitrack REMI encoding, and given the MIR models we have, 
it incorporates 
control tokens such as genre, chord, and section tokens to the encoding.

\subsection{Controllable Symbolic Music Generation}
Previous methods for controlling symbolic music generation have typically fallen into two categories.
The first is based on Variational Autoencoders (VAEs) \cite{roberts2018hierarchical,wu2023musemorphose}.
VAEs aim to find a latent space for representing music that encodes distinct musical attributes in independent dimensions. This disentanglement allows for specific attributes of generated music (e.g., rhythm, genre, or timbre) to be individually manipulated by altering corresponding dimensions in the latent space without affecting other attributes, thereby enhancing the controllability of music generation.

The second approach is to include control tokens in the encoding of symbolic music. For example, MMM \cite{ens2020mmm} includes instruments and note density tokens in the encoding, which can be specified at inference. Similarly, FIGARO \cite{von2022figaro} uses ``expert descriptions'' 
indicating time signature, note density, mean pitch, mean velocity and mean duration as well as instruments and chords.
It then uses an encoder-decoder model to learn a mapping from descriptions to sequences of a piece of music.
Driven by the development of Large Language Models (LLMs), recent work has also explored using natural language to control symbolic music generation \cite{sheng2021songmass,hussain2023m,lu2023musecoco,yuan2024chatmusician}. Natural language text can also be treated as control tokens, with the key distinction that it usually requires pre-training the LLM on text.

In our work, the proposed SymPAC framework is designed to work with a decoder-only language model.
In a controlled generation setting, prompt bars that conform with user input control signals are generated first. The generation of musical part comes after that, in which the model will have full context of control signals from prompt bars. These two generation stages are both controlled by an FSM, which takes into consideration the grammar of the encoding and user inputs.
There are two main
differences between SymPAC and previous works
\begin{enumerate}
\item We encode control signals as tokens and use FSM to enforce input control signals during inference.
In contrast, for VAE-based control methods, control signals are converted into latent embeddings, and the model is not guaranteed to follow these control signals.
\item Since we use a decoder-only language model, the tokens in prompt bars are also learned simultaneously.
Consequently, the user is only required to input a portion of the control information, with the model being able to automatically generate missing controls.
In contrast, an encoder-decoder framework like the one described in \cite{von2022figaro} would require a complete encoder input during inference, which lacks flexibility.
\end{enumerate}

\section{Method}\label{sec:method}

\begin{figure*}[!t]
\centering
\includegraphics[width=0.9\linewidth]{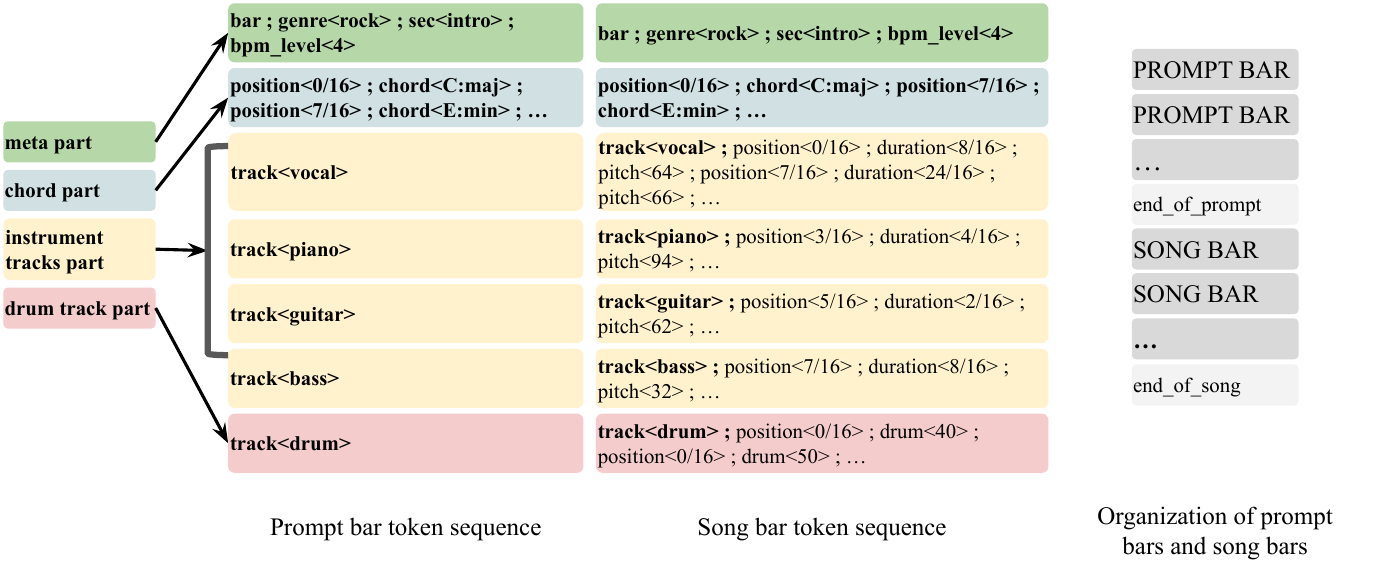}
\caption{Illustration of our symbolic music encoding.}
\label{fig:encoding}
\end{figure*}

\subsection{Symbolic Music Encoding And Prompt Bars}\label{sec:representation}
Our data representation is based on the REMI+ \cite{von2022figaro} representation, an extension of REMI \cite{huang2020pop} that supports multitrack data.
An illustration of our encoding is shown in Fig.~\ref{fig:encoding}.
The fundamental unit of our encoding is a bar, of which there are two types: \textit{prompt bar} and \textit{song bar}.
The token sequence of a song bar can be divided into four parts:

\begin{itemize}
\item The \textit{meta} part includes four tokens for the \texttt{bar}, \texttt{genre}, \texttt{sec} (for section type name), and \texttt{bpm\_level} (which indicates the tempo range).
\item The \textit{chord} part consists of alternating \texttt{position} and \texttt{chord} tokens.
\item Each \textit{instrument track} part consists of a \texttt{track} token, followed by one or more groups of \texttt{position}, \texttt{duration} and \texttt{pitch} tokens.
\item The \textit{drum track} part consists of a \texttt{track<drum>} token, 
followed by one or more groups of \texttt{position} and \texttt{drum} (drum MIDI) tokens.
\end{itemize}

Here are further explanations of \texttt{position}, \texttt{duration} and \texttt{track} tokens\footnote{Details of all token types are provided in supplementary materials}:

\begin{itemize}
\item \texttt{position}: Represents the starting position of subsequent \texttt{chord}, \texttt{pitch} or \texttt{drum} token within a bar. Each bar is divided into 16 steps, so that position ranges from 0/16 to 15/16.
\item \texttt{duration}: Ranges from the minimum time division of 1/16 bar to a maximum of 2 bars, or 32/16.
\item \texttt{track}: A track token will only exist if there is at least one note in the bar for the corresponding instrument.
This allows the user to control which instruments are used within a bar.
\end{itemize}

Prompt bars contain a subset of tokens in song bars, retaining only tokens that represent \textit{control signals}.
In our case, these include genre, section, tempo, chords and tracks.
As future work, this encoding could be extended to include more control signals (e.g. note density for a track).
The encoding of a full piece of music will consist of: all prompt bars in the piece; then, a special \texttt{end\_of\_prompt} token; then, all song bars in the piece; and finally a special \texttt{end\_of\_song} token.

During training stage, the model is trained to predict tokens in prompt bars as well,
not distinguishing them from
tokens in song bars.
As mentioned previously, this design enables the user to input partial control signals (or no input at all), and the model is able to infer the missing ones.

\begin{algorithm}
\caption{Constrained Generation via FSM}
\label{alg:fsm}
\begin{algorithmic}[1]
\Procedure{ConstrainedSampling}{$\mathcal{M}$, $\mathcal{V}$, $\mathcal{R}$}
    \State $\mathbf{s}_0 \gets$ $x_0$ start token (\texttt{bar} in our encoding)
    \State $q_0 \gets$ initial state
    \State $t \gets 0$
    \While{not end of sequence}
        \State $\mathcal{V}_{t+1} \gets \textsc{GetSubVocab}(\mathcal{R}, q_t, x_t)$
        \State $q_{t+1} \gets \textsc{UpdateState}(\mathcal{R}, q_t, x_t)$
        \State $x_{t+1} \gets \textsc{Sample}(\mathcal{M}, \mathcal{V}_{t+1})$
        \State $\mathbf{s}_{t+1} \gets \mathbf{s}_t \circ x_{t+1}$
        \State $t \gets t + 1$
    \EndWhile
    \State \textbf{return} $\mathbf{s}_{t}$
\EndProcedure
\end{algorithmic}
\end{algorithm}

\subsection{Constrained Generation via FSM}
In the controlled symbolic music generation setting, there are two types of constraints:

\noindent\textbf{Grammar constraint}: The encoding of symbolic music follows a specific format. For example, for our proposed encoding shown in Fig.~\ref{fig:encoding}, a \texttt{bar} token will always be followed by a \texttt{genre} token.

\noindent\textbf{User input constraint}: 
Generated token sequence should conform with user inputs.
For example, if the user wants to generate ``rock'' style music, the \texttt{genre} token can only be \texttt{genre<rock>}.

Since we are already aware of these constraints in advance,
there is no need to sample from the entire vocabulary space during inference.
Instead, we can sample from a subspace that is in accordance with the constraints.

To achieve this, we employ a Finite State Machine (FSM) to interact with the language model $\mathcal{M}$ during inference. Let $x_{t}$ denote the token generated by $\mathcal{M}$ at time step $t$. The FSM takes $x_{t}$, the current state $q_{t}$ and the predetermined rule set $\mathcal{R}$, and outputs a subset of the vocabulary $\mathcal{V}_{t+1}$, from which the language model $\mathcal{M}$ can sample at time $t+1$. We call this procedure Constrained Generation via FSM, which is formally defined in Algorithm \ref{alg:fsm}.
This algorithm is analogous to regular expression matching, where it checks if a given input string conforms to a specified pattern.
Here the pattern and input string are equivalent to rule set $\mathcal{R}$ and token sequence $s_t$ respectively.

\section{Experiments And Results}
\label{sec:experiment}

To validate our contributions, we conduct experiments to assess whether the system is scalable (i.e., improves when scaling up training data) and controllable (i.e., there is consistency between generation output and user inputs).

In Sec.~\ref{sec:uncond}, we conduct a quantitative analysis to compare models trained on different amounts of training data, in order to assess scalability.
In Sec.~\ref{sec:cond}, we examine two common types of control inputs: chord progression and section structure.
The impact of these control inputs is tested through both quantitative metrics and qualitative examples.
Lastly, in Sec.~\ref{subsec_subjective}, we compare our models trained on different datasets with other baseline symbolic music generation systems through subjective evaluation.

\subsection{Datasets}
We use three datasets in our experiments. We always use each dataset individually; i.e., we never merge the datasets to train a single model. The datasets are:

\vspace{4pt}

\noindent \textbf{Lakh MIDI Dataset (LMD)} \cite{lakhmidi}. A dataset in MIDI format, containing around 170K songs.
We use this to compare with models trained on transcribed audio data.

\noindent \textbf{Million Song Dataset (MSD)} \cite{msd}. A public dataset used extensively by MIR researchers. We use the 
30--60s preview audio clips, representing the highlight of the song.

\noindent \textbf{In-House Dataset (IHD)}. We use a licensed internal collection with about 1M 
full songs in audio format, covering a wide range of Western modern genres.

\subsection{Training Settings}
We train a decoder language model with the Llama \cite{touvron2023llama} architecture. We set the number of layers, number of attention heads and embedding dimensions to be 12, 12 and 768 respectively, resulting in a model with about 86M trainable parameters.
We concatenate token sequences of all pieces into a 1-D array, and randomly pick a window of size 10,240 as one training sample. As the average sequence lengths of LMD, MSD and IHD are 900, 1500 and 8000 respectively, this window size would contain 11.4, 6.8 and 1.3 pieces on average for each dataset.

When training data are limited, data augmentation and data filtering (to ensure that unusual data do not pollute the training) are commonly used.
However, we adopt neither approach, for two reasons.
First, since we have a large dataset of audio samples,  the training data are likely to cover a broad spectrum of examples already, reducing the need to filter out unusual data points.
Second,  augmentation may alter the training data in unwanted ways.
For example, a common augmentation approach is to transpose all the pitches in a piece \cite{huang2018music,dong2023multitrack}.
However, this may distort the pitch ranges of each instrument: e.g., if the input bass parts are transposed up and down, the model will not learn the correct range of realistic bass notes.

\begin{table}[!ht]
 \begin{center}
 \begin{tabular}{cccc}
  \hline
  \textbf{Metric Class} & \textbf{IHD 100\%} & \textbf{IHD 10\%} & \textbf{IHD 1\%}\\
  \hline\hline
  Chord & \textbf{0.112} & 0.119 & 0.347 \\
  Structure & 0.348 & \textbf{0.220} & 0.786 \\
  Vocal Note & \textbf{0.416} & 0.892 & 1.086 \\
  Guitar Note & \textbf{0.222} & 0.257 & 0.397 \\
  Piano Note & \textbf{0.178} & 0.403 & 0.686 \\
  Bass Note & \textbf{0.180} & 0.867 & 1.038 \\
  Drum Note & \textbf{0.650} & 2.902 & 1.248 \\
  \hline
 \end{tabular}
 \end{center}
 \centering
\caption{Average Kullback-Leibler Divergence (KLD) of metrics in different metric classes for models trained on different dataset against a held-out validation set.}
 \label{tab:uncond}
\end{table} 

\subsection{Unconditioned Generation}
\label{sec:uncond}
Intuitively, increasing the amount of data should enhance the performance of the model.
In this experiment, we use objective metrics to validate this.
Designing objective metrics to evaluate symbolic music remains an open question.
A common approach is to prepare a reference dataset,
calculate embeddings or metrics
of the generated samples and reference set,
and then compare these using distance metrics such as the Fréchet Distance or Kullback-Leibler Divergence (KLD). 
For a detailed review on evaluation methods for symbolic music, see \cite{yang2020evaluation,ji2023survey}.

In our experiment, we prepare a held-out validation set with 3000 samples.
We use a range of metrics that can be categorized into the following classes: chord, structure, instrument note (including vocals, guitar, piano and bass) and drum note.
Detailed definitions are provided in supplements. In general, the metrics in each class are as follows:
\begin{itemize}
\item \textbf{Chord}: chord label, chord root, chord transition;
\item \textbf{Structure}: section label, section label bigram, instrument labels per bar;
\item \textbf{Instrument Note}: note pitch, note duration, pitch class, min/max pitch per bar, max number of notes per bar, uniformity of number of notes per bar;
\item \textbf{Drum Note}: drum key, max number of notes per bar, uniformity of number of notes per bar, and unique drums per bar.
\end{itemize}

We compare models trained on  100\%, 10\% and 1\% of the IHD data,
and do generation in an unconditioned setting.
For each model, we generate 800 samples to compute metric distributions.
KLD values are then computed between distributions of generated samples and distribution of the validation set for each metric. 
Lower KLD indicates that two distributions are closer, suggesting the generated samples sound more similar to the validation set.
We report the average KLD values for the same class, and provide a full list of KLDs for each metric in supplements.

The results are shown in Table \ref{tab:uncond}. We can see that the model trained with 100\% IHD data has the lowest KLD against the validation set on 6 out of 7 classes, and the model trained on only 1\% data has the highest KLD on 6 out of 7 classes.
The results confirm that a model trained on more data can generate samples closer to the training data.
Furthermore, we observed that
the benefit of using more data is greater for the 'Note' metrics than for the 'Chord' or 'Structure' ones.
This is likely because note tokens are more numerous and have complex distributions, which needs larger scale of data to learn.
Counterintuitively, the KLD for 'Structure' was better when using 10\% of the data instead of 100\%. We speculate that since the structure tokens are scarcest, this could be the result of a lucky alignment between the validation set at the 10\% of the data used, but this deserves more study.

\begin{table*}[!ht]
 \begin{center}
\begin{tabular}{llllll}
\hline
 \textbf{Model} & \textbf{Coherence} & \textbf{Richness} & \textbf{Arrangement} & \textbf{Structure} & \textbf{Overall} \\
 \hline\hline
FIGARO & 3.12 ± 0.82 & 2.73 ± 0.92 & 2.85 ± 0.96 & 2.62 ± 0.80 & 2.74 ± 0.89 \\
MMT & 2.37 ± 0.35 & 2.27 ± 0.36 & 2.37 ± 0.34 & 2.08 ± 0.30 & 2.16 ± 0.35 \\
Ours (IHD) & \textbf{3.55 ± 0.53} & \textbf{3.58 ± 0.38} & \textbf{3.45 ± 0.49} & \textbf{3.73 ± 0.32} & \textbf{3.60 ± 0.39} \\
Ours (LMD) & 3.25 ± 0.34 & 3.25 ± 0.35 & 3.28 ± 0.30 & 3.20 ± 0.61 & 3.25 ± 0.46 \\
Ours (MSD) & 3.16 ± 0.27 & 3.17 ± 0.33 & 3.09 ± 0.32 & 3.15 ± 0.29 & 3.07 ± 0.28 \\
\hline
\end{tabular}
 \end{center}
 \centering
\caption{Results of subjective evaluation, mean opinion score (MOS)}
 \label{tab:mos}
\end{table*}

\subsection{Controlled Generation}
\label{sec:cond}
The SymPAC framework aims to give users flexible control over the music generation process. However, we need to verify that this control is effective: do the notes generated agree with the control inputs?
To this end, we conduct controlled generation experiments on two input scenarios: chord progression inputs and section structure inputs.

\noindent \textbf{Chord Progression Inputs}.
In this experiment, we randomly pick 20 top trending chord progressions from \textit{HookTheory}\footnote{https://www.hooktheory.com} as the chord progression inputs.
We only include major and minor triad chords.
We then let the model generate 64 bars of music by looping the chord progressions.
To evaluate the match between the input chord progression and the output, we apply a symbolic chord detection method on the generated samples. Details about the method can be referred in the supplementary materials.

The accuracy of detected chord from the input chord progression is shown in Table \ref{tab:chord}.
As shown in the result, the models trained on MSD, IHD 100\% and IHD 10\% all have similar overall accuracy, with MSD slightly out-performing the others. But the model trained on IHD 1\% (just 10K songs) is much worse than the other three. This suggests that a dataset at the scale of 100K songs is sufficient to model low-level control signals like chord, given the model and encoding we are using here.
We also provide examples in supplementary audios of outputs when given unusual chord progressions.

\begin{table}[!ht]
 \begin{center}
 \begin{tabular}{cc}
  \hline
  \textbf{Training Dataset} & \textbf{Accuracy} \\
  \hline\hline
  IHD 100\% & 87.2\% \\
  IHD 10\% & 86.9\% \\
  IHD 1\% & 74.0\% \\
  MSD & \textbf{87.6\%} \\
  \hline
 \end{tabular}
 \end{center}
 \centering
\caption{Accuracy of chord progressions in controlled generation with chord input.}
 \label{tab:chord}
\end{table}

\noindent \textbf{Section Structure Inputs}.
In this experiment, we take 10 typical section sequences as inputs (listed in supplements), ranging in length from 4 to 13 sections (16 to 68 bars), and use each model to generate 100 outputs per prompt.
We compare the same 4 models from the previous section.
For each generated output,
we leverage a Music Structure Analysis (MSA) algorithm~\cite{nieto2020audio-based} to predict its structure, and compare this to the input structure.
The MSA algorithm's predictions may be inaccurate, but we still expect that a greater match between the intended and estimated structure indicates more success at controlling the structure.
We use Foote's algorithm~\cite{foote2000automatic} for segmentation and the 2D-Fourier magnitude algorithm~\cite{nieto2014music}
for section labeling, with a beat-wise feature embedding that averages the pitch-wise MIDI piano rolls within a beat interval. We evaluate the results using \texttt{mir\_eval}~\cite{raffel2014}, and report three metrics: boundary prediction \textit{f}-measure with a 3-second tolerance (HR3F); pairwise clustering \textit{f}-measure (PWF); and the normalized entropy score \textit{f}-measure (Sf). To test directly how similar the repeated sections are, we also report PWF and Sf when the ground-truth segmentation is used.

We find that all metrics are worse (lower) when the system is trained on MSD or IHD 1\%, and improve substantially when at least 10\% of the data are used (Table~\ref{tab:structure}). This is expected, since the audio clips in MSD are only excerpts and thus not instructive for modelling full-song structure.

\begin{table}[!ht]
 \begin{center}
 \begin{tabular}{l|ccc|cc}
  \hline
  & \multicolumn{3}{c|}{\textbf{Regular}} & \multicolumn{2}{c}{\textbf{Oracle}} \\
\textbf{Dataset} & \textbf{HR3F} & \textbf{PWF} & \textbf{Sf} & \textbf{PWF} & \textbf{Sf} \\
  \hline\hline
  IHD 100\%  & \textbf{0.60} & \textbf{0.50} & \textbf{0.50} & \textbf{0.72} & \textbf{0.80} \\
  IHD 10\%   & \textbf{0.60} & 0.49 & 0.49 & 0.70 & 0.79 \\
  IHD 1\%    & 0.54 & 0.47 & 0.47 & 0.62 & 0.73 \\
  MSD        & 0.57 & 0.47 & 0.47 & 0.63 & 0.74 \\

  \hline
 \end{tabular}
 \end{center}
 \centering
\caption{Accuracy of structure predicted from generated songs with no guidance (left) and with ground truth segmentation (right).}
 \label{tab:structure}
\end{table}

\begin{figure}
 \centerline{
 \includegraphics[width=0.8\columnwidth]{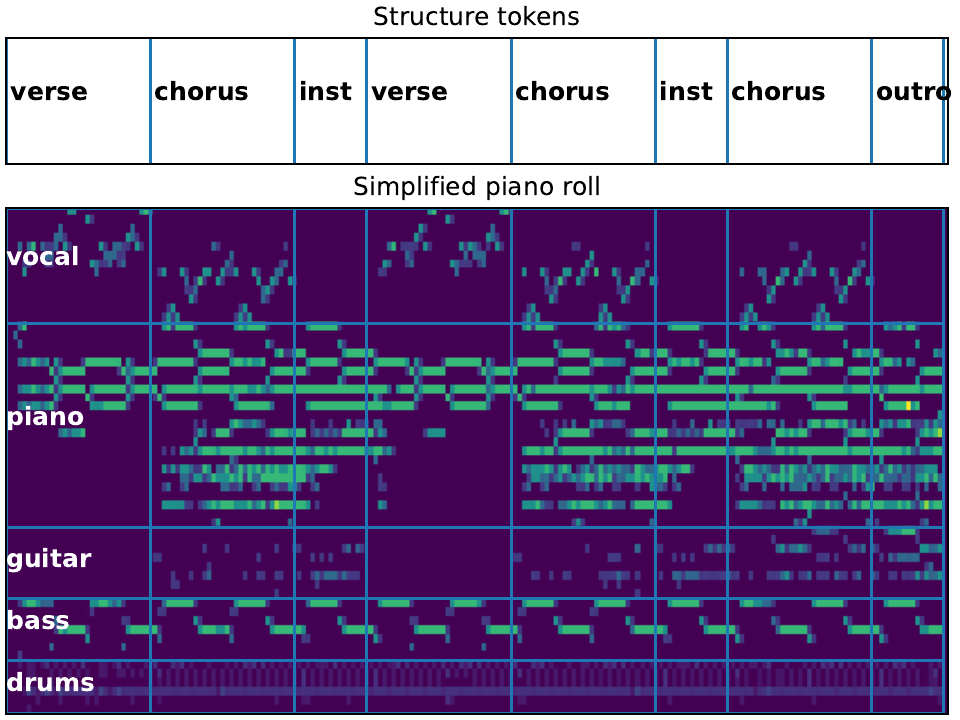}}
 \caption{Constrained generation output with user-defined structure using IHD 100\% model. The simplified piano roll gives beat-averaged values and excludes empty lines.}
 \label{fig:example_structure}
\end{figure}

Fig.~\ref{fig:example_structure} shows the piano roll 
of a typical output, where the match between the intended and predicted structure was average (Sf = 0.508).
Even so,
the match between the intended and realized structure is evident in the piano roll: the chorus sections are similar but not identical to each other, and so are the verse sections.

In both controlled generation experiments, the gap between 100\% and 10\% IHD is very small, indicating that 10\% IHD data combined with SymPAC is sufficient for achieving good adherence to control inputs. However, it is important to remember that the metrics of these two experiments only reflect whether the control signals are well-followed, not the overall quality of the generated pieces.

\subsection{Subjective Evaluation}
\label{subsec_subjective}
The models tested so far were all trained on transcribed audio data, so it is worth comparing
with models trained directly on MIDI data.
In this experiment, we compare our model trained on LMD, MSD and IHD, and also two baselines, FIGARO \cite{von2022figaro} and MMT \cite{dong2023multitrack}, in a subjective listening test.
We recruited 12 participants with the background of MIR researchers or music producers.
Similar to \cite{dong2023multitrack}, we asked each participant to rate 10 audio samples generated by each model on a 5-point Likert scale on five criteria: coherence, richness, arrangement, structure and overall~\footnote{These criteria are described as: (1) Coherence: The rhythm is stable; The chord progression develops logically; Dissonant notes are not excessive. (2) Richness: The melody and acccompaniment are interesting and diverse.
(3) Arrangement: Collaboration among multiple instruments is harmonious and natural; Arrangements for different instruments are diverse and reasonable.
(4) Structure: The piece includes a clear and engaging structure with appropriate repetitions and variations; The piece has obvious connections and reasonable developments between sections.
(5) Overall: I like this piece in general.
}.

The result is summarized in Table \ref{tab:mos}. All of our 
proposed models outperform the baselines in all dimensions.
Our model trained on IHD has higher performance than the other two training data setups,
which attests to the viability of leveraging audio data by running MIR models at scale.
The result using LMD was better than MSD, despite having fewer songs;
this could be due to LSD having more notes than MSD (see Tab~\ref{tab:dataset}), or due to it containing full songs instead of only excerpts.
We only compare FIGARO and Ours (LMD) with a statistical test, since these were trained on the same dataset. Mann-Whitney U tests found significant differences in Richness (p = .005), Structure (p = .0005), and Overall (p = .027) ratings, but not in Coherence (p = .85) or Arrangement (p = .122).

\section{Conclusions And Future Work}
We trained a language model for symbolic music generation leveraging audio data and pre-trained MIR models.
We proposed the SymPAC framework, which includes prompt bars in encoding and Constrained Generation via FSM during inference time.
We showed how  combining these two components enables a user to control the generation process, and we evaluated the results through quantitative and qualitative analysis.

Future work could improve at least two aspects of this system:
(1) We quantified position and duration to 1/16 per bar, which does not support 3/4 or 6/8 time signatures well. Also, the chord detection model we used only supports 12 major and minor chords, limiting the user input options. We can expand the encoding to support finer-grained quantization and more advanced chords.
(2) Our token sequence length is long: 8000 on average for samples in IHD.
We could use tokenization methods such as Byte Pair Encoding \cite{fradet2023byte} or use compound word tokens \cite{hsiao2021compound} to compress sequences and improve training efficiency.

\bibliography{ISMIRtemplate}

%
%
%
%
%
\clearpage

\subfile{supplements}

\end{document}

%% file: supplements.tex
\twocolumn[
\begin{center}
    \Large \textbf{\MakeUppercase{Supplementary Materials}}
\end{center}
\vspace{1cm}  
]
\setcounter{section}{0}
\setcounter{figure}{0}
\setcounter{table}{0}
\section{Details For MIR Models}

We employ specialized MIR models, including beat tracking \cite{hung2022modeling}, chord recognition \cite{lu2021spectnt}, structure segmentation \cite{wang2022catch, wang2022musfa}, 
vocal melody transcription \cite{lu2023multitrack}, 
and multi-instrument transcription \cite{lu2023multitrack}. These models leverage Transformer-based architecture and have undergone state-of-the-art performance testing on respective benchmark tasks. Below are the specifics regarding the output of each model:
\begin{itemize}
\item \emph{Beat}: Each beat is characterized by its timestamp and beat count label, where count = 1 indicates a downbeat. The model supports time signatures of 3/4 and 4/4.
\item \emph{Chord}: For each chord, the starting and ending times, along with the chord name, are recorded. The chord names cover the 12 major and minor chords, along with a `none' designation.
\item \emph{Section}: Sections are identified by their starting and ending times, accompanied by function labels covering `intro', `verse', `chorus', `instrumental', `bridge', `outro', and `silence'.
\item \emph{Vocal Melody}: The model outputs a list of non-overlapping notes, each comprising onset and offset times along with a pitch key, assuming the melody is monophonic. The pitch keys follow the General MIDI key map.
\item \emph{Instrument Transcription}: Four instruments are transcribed: bass, guitar, piano, and drums. Each instrument is associated with a list of notes, each containing onset and offset times along with a pitch key. Note overlaps are possible. The pitch keys adhere to the General MIDI key and percussion map.
\end{itemize}

\section{Details For Encoding}
Our encoding contains the types of tokens defined in Table \ref{tab:tokens}.

\section{Details For Chord Control Experiment}
Table \ref{tab:chords} shows the chord progressions we used in the chord controlled generation experiment.

\begin{table}[!ht]
\begin{center}
\begin{tabular}{l}
\hline
\textbf{Chord progressions} \\
\hline\hline
F, D:min, C, G \\
F, C, G, A:min \\
C, A:min, E:min, F \\
C, A:min, F, G \\
A:min, G, F, G \\
A:min, F, C, G \\
A:min, G, F, E:min, D:min, G, C, C \\
F, G, E:min, A:min, D:min, G, C, C \\
C, G, A:min, E:min, F, C, D:min, G \\
C, G, A:min, F, C, G, E:min, F \\
F, D:min, E, A:min \\
A:min, G, F, E \\
C, F, A\#, C \\
G, E:min, C, D \\
C, G:min, A\#, D:min \\
C, G\#, F, G \\
A:min, G, F, F:min \\
C, A, D, G \\
A:min, G:min, F, E \\
A:min, D:min, G, E, F, D:min, E, A:min \\
\hline
\end{tabular}
\end{center}
\centering
\caption{Chord progressions used in chord controlled generation experiment.}
\label{tab:chords}
\end{table}

To detect the chord on the generated samples, we apply a chroma-based method. For each chord symbol, we learn a 12-d chroma template to represent pitch class distribution based on statistics on the POP909 dataset \cite{wang2020pop909}. Then on each bar of the generated sample, we calculate a chroma vector and detect chord by auto-correlation on the chord chroma templates. This method achieved $97.6\%$ accuracy on the test set.

\section{Details For Uncondition Generation Experiment}
The full list of metrics and their meanings is listed in Table~\ref{tab:metric}. The Kullback-Leibler Divergence (KLD) values for each metric is shown in Table~\ref{tab:kld}.

\section{Details For Structure Control Experiment}
Table \ref{tab:struct} shows the structure inputs we used in the structure controlled generation experiment.

\begin{table*}[!ht]
\begin{center}
\renewcommand{\arraystretch}{1.5}
\begin{tabular}{cM{8cm}}
\hline
\textbf{Index} & \textbf{Structure} \\
\hline\hline
    1 & intro * 4, verse * 8, chorus * 8, outro * 4 \\
    2 & intro * 4, verse * 8, chorus * 8, inst * 8, chorus * 8 \\
    3 & intro * 4, verse * 8, chorus * 8, verse * 8, chorus * 8 \\
    4 & intro * 4, verse * 8, chorus * 8, inst * 8, verse * 8, chorus * 8, outro * 4 \\
    5 & intro * 4, verse * 4, chorus * 8, inst * 4, verse * 4, chorus * 8, inst * 4, chorus * 8, outro * 4 \\
    6 & intro * 4, verse * 8, chorus * 8, inst * 8, verse * 8, chorus * 8, bridge * 8, chorus * 8, outro * 4 \\
    7 & intro * 4, verse * 8, chorus * 8, inst * 8, verse * 8, chorus * 8, inst * 4, bridge * 4, chorus * 8, outro * 4 \\
    8 & verse * 8, chorus * 8, inst * 4, verse * 8, chorus * 8, inst * 4, chorus * 8, outro * 4 \\
    9 & chorus * 8, verse * 8, chorus * 8, inst * 8, verse * 8, chorus * 8, outro * 4 \\
    10 & intro * 4, chorus * 8, verse * 8, inst * 4, verse * 4, inst * 4, chorus * 8, verse * 4, inst * 4, verse * 4, bridge * 4, chorus * 8, outro * 4 \\
\hline
\end{tabular}
\end{center}
\centering
\caption{Chord progressions used in chord controlled generation experiment.}
\label{tab:struct}
\end{table*}

\begin{table*}[!ht]
\renewcommand{\arraystretch}{1.5}
\begin{center}
\begin{tabular}{cM{12cm}}
\hline
\textbf{type} & \textbf{options for} \texttt{X} \\
\hline\hline
\texttt{bar} & N/A \\
\texttt{end\_of\_prompt} & N/A \\
\texttt{sec<X>} & silence, intro, verse, chorus, inst, bridge, outro \\
\texttt{bpm\_level<X>} & <82, [82,96), [96,110), [110,120), [120,125), [125,132), [132,143), >=143 \\
\texttt{genre<X>} & Blues, Childhood, Classical, Country, Easy\_Listening, Electronic, Experimental, Folk, Hip\_Hop/Rap, Jazz, Latin, Metal, New\_Age, Pop, R\&B/Soul, Reggae, Rock \\
\texttt{position<X/16>} & 0, 1, 2, ..., 15 \\
\texttt{chord<X>} & C:maj, C\#:maj, D:maj, D\#:maj, E:maj, F:maj, F\#:maj, G:maj, G\#:maj, A:maj, A\#:maj, B:maj, C:min, C\#:min, D:min, D\#:min, E:min, F:min, F\#:min, G:min, G\#:min, A:min, A\#:min, B:min, C:sus2, C\#:sus2, D:sus2, D\#:sus2, E:sus2, F:sus2, F\#:sus2, G:sus2, G\#:sus2, A:sus2, A\#:sus2, B:sus2, C:sus4, C\#:sus4, D:sus4, D\#:sus4, E:sus4, F:sus4, F\#:sus4, G:sus4, G\#:sus4, A:sus4, A\#:sus4, B:sus4, C:aug, C\#:aug, D:aug, D\#:aug, E:aug, F:aug, F\#:aug, G:aug, G\#:aug, A:aug, A\#:aug, B:aug, C:dim, C\#:dim, D:dim, D\#:dim, E:dim, F:dim, F\#:dim, G:dim, G\#:dim, A:dim, A\#:dim, B:dim, N \\
\texttt{track<X>} & vocal, piano, guitar, bass, drums \\
\texttt{pitch<X>} & 0, 1, 2, ..., 127 \\
\texttt{duration<X/16>} & 1, 2, 3, ..., 32 \\
\texttt{drum<X>} & 35, 36, 37, ..., 81 \\
\hline
\end{tabular}
\end{center}
\centering
\caption{Token types and options.}
\label{tab:tokens}
\end{table*}

\begin{table*}[!ht]
 \begin{center}
 \begin{tabular}{lll}
  \hline
  \textbf{Type} & \textbf{Metric Name} & \textbf{Definition} \\
  \hline\hline
  \multirow{3}{*}{Chord} & Chord Label & The distribution of chord label (1 chord for each beat, same below). \\
  & Chord Root & The distribution of chord root. \\
  & Chord Transition & The distribution of the tuple of 2 consecutive and different chords. \\
  \hline
  \multirow{3}{*}{Structure} & Section Label & The distribution of section labels, weighted by number of bars of each section. \\
  & Section Label Bigram & The distribution of section label bigram (2 consequtive section) \\
  & Instrument Labels per bar & The distribution of a list of instruments in a bar. \\
  \hline
  \multirow{7}{*}{Inst Note~*} & Note Pitch & The distribution of note pitches. \\
  & Note Duration & The distribution of note durations. \\
  & Pitch Class & The distribution of note pitch class. \\
  & Min Pitch Per Bar & The distribution of minimum pitch values for each bar. \\
  & Max Pitch Per Bar & The distribution of maximum pitch values for each bar. \\
  & Max \#Notes & The maximum number of notes per bar. \\
  & \#Notes Uniformity & The relative entropy of number of notes per bar v.s. uniform distribution. \\
  \hline
  \multirow{4}{*}{Drum Note} & Drum Pitch & The distribution of drum pitches. \\
  & Max \#Drum & The maximum number of drum notes per bar. \\
  & \#Drum Uniformity & The relative entropy of number of drum notes per bar v.s. uniform distribution. \\
  & Unique Drums Per Bar & The distribution of a list of unique drum notes for each bar. \\
  \hline
 \end{tabular}
\end{center}
\centering
 \caption{Metrics we use to compare distribution of generated samples against a held-out validation dataset. *~Inst Note includes vocal, guitar, piano and bass notes.}
 \label{tab:metric}
\end{table*}

\begin{table*}[!ht]
 \begin{center}
 \begin{tabular}{lllll}
  \hline
  \textbf{Type} & \textbf{Metric Name} & \textbf{IHD 100\%} & \textbf{IHD 10\%} & \textbf{IHD 1\%} \\
  \hline\hline
\multirow{3}{*}{Chord} & Chord Label & 0.151 &    0.116 &   0.338  \\
& Chord Root & 0.063 &    0.063 &   0.181  \\
& Chord Transition & 0.121 &    0.178 &   0.521  \\
\hline
\multirow{3}{*}{Structure} & Section Label & 0.099 &    0.032 &   0.333  \\
& Section Label 2-Gram & 0.883 &    0.500 &   1.712  \\
& Instrument Labels Per Bar & 0.061 &    0.127 &   0.313  \\
\hline
\multirow{7}{*}{Vocal Note} & Note Pitch & 0.103 &    0.162 &   0.325  \\
& Note Duration & 0.126 &    0.167 &   0.261  \\
& Pitch Class & 0.010 &    0.025 &   0.174  \\
& Min Pitch Per Bar & 0.125 &    0.141 &   0.326  \\
& Max Pitch Per Bar & 0.159 &    0.246 &   0.325  \\
& Max \#Notes & 2.222 &    5.341 &   5.345  \\
& \#Notes Uniformity & 0.164 &    0.164 &   0.843  \\
\hline
\multirow{7}{*}{Guitar Note} & Note Pitch & 0.063 &    0.090 &   0.340  \\
& Note Duration & 0.153 &    0.118 &   0.353  \\
& Pitch Class & 0.018 &    0.028 &   0.127  \\
& Min Pitch Per Bar & 0.197 &    0.127 &   0.478  \\
& Max Pitch Per Bar & 0.145 &    0.194 &   0.377  \\
& Max \#Notes & 0.350 &    0.485 &   0.659  \\
& \#Notes Uniformity & 0.630 &    0.758 &   0.444  \\
\hline
\multirow{7}{*}{Piano Note} & Note Pitch & 0.056 &    0.089 &   1.034  \\
& Note Duration & 0.083 &    0.068 &   0.430  \\
& Pitch Class & 0.014 &    0.020 &   0.183  \\
& Min Pitch Per Bar & 0.112 &    0.200 &   0.900  \\
& Max Pitch Per Bar & 0.123 &    0.171 &   1.000  \\
& Max \#Notes & 0.211 &    1.346 &   0.644  \\
& \#Notes Uniformity & 0.645 &    0.924 &   0.611  \\
\hline
\multirow{7}{*}{Bass Note}& Note Pitch & 0.136 &    0.094 &   0.391  \\
& Note Duration & 0.060 &    0.076 &   0.154  \\
& Pitch Class & 0.026 &    0.038 &   0.294  \\
& Min Pitch Per Bar & 0.147 &    0.165 &   0.325  \\
& Max Pitch Per Bar & 0.175 &    0.267 &   0.417  \\
& Max \#Notes & 0.439 &    4.893 &   4.844  \\
& \#Notes Uniformity & 0.278 &    0.539 &   0.843  \\
\hline
\multirow{4}{*}{Drum Note} & Drum Pitch & 0.124 &    0.251 &   0.257  \\
& Max \#Drum & 0.958 &    8.696 &   2.552  \\
& \#Drum Uniformity & 0.668 &    0.851 &   0.688  \\
& Unique Drums Per Bar & 0.848 &    1.811 &   1.495  \\
  \hline
 \end{tabular}
\end{center}
\centering
 \caption{Kullback-Leibler Divergence (KLD) of each metric for samples generated from IHD 100\%, 10\% and 1\% trained models against the same held-out validation set.}
 \label{tab:kld}
\end{table*}
